\documentclass[conference]{IEEEtran}
\IEEEoverridecommandlockouts
\usepackage{cite}
\usepackage{amsmath,amssymb,amsfonts}
\usepackage{algorithmic}
\usepackage{graphicx}
\usepackage{textcomp}
\usepackage{enumitem}

\usepackage{xcolor}
\usepackage[T1]{fontenc}
\usepackage{bm}
\usepackage{caption}
\usepackage{times} 
\DeclareCaptionFont{eightpt}{\fontsize{8pt}{9.5pt}\selectfont}
\usepackage{multirow}
\definecolor{ours}{RGB}{0,102,204}      
\definecolor{reported}{RGB}{120,120,120} 
\definecolor{reranker}{RGB}{0,153,76}    

\usepackage{float}
\usepackage{tikz}
\usepackage{url}
\usepackage{pgfplots}

\usepackage[table]{xcolor}
\usepackage[switch]{lineno}
\usepackage{booktabs}
\usepackage[dvipsnames]{xcolor} 
\definecolor{myLightOrange}{RGB}{255, 204, 153}
\definecolor{myLightPurple}{RGB}{221, 160, 221}
\definecolor{myLightBlue}{RGB}{224, 236, 244} %
\definecolor{myLightGreen}{RGB}{204, 235, 197}  
\definecolor{myLightBlue}{RGB}{198, 219, 239}   
\definecolor{myLightYellow}{RGB}{255, 255, 224} 
\usepackage{ragged2e} 

\definecolor{lightyellow}{RGB}{255, 255, 200}

\usepackage{makecell}

\usepackage{array}

\pgfplotsset{compat=1.18} 
\usepackage{subcaption}
\def\BibTeX{{\rm B\kern-.05em{\sc i\kern-.025em b}\kern-.08em
    T\kern-.1667em\lower.7ex\hbox{E}\kern-.125emX}}
\begin{document}




\title{
\textit{ScaleCall} \\ Agentic Tool Calling at Scale for Fintech: Challenges, Methods, and Deployment Insights
}



\author{
\IEEEauthorblockN{$^{1,3}$Richard Osuagwu$^\ast$\thanks{*Work done during internship at Mastercard, Ireland}, $^{2,3}$Thomas Cook$^\ast$, $^{3}$Maraim Masoud, $^{3}$Koustav Ghosal, $^3$Riccardo Mattivi}
\\
\IEEEauthorblockA{
$^1$Maynooth University, Ireland\\
$^2$TU Dublin, Dublin, Ireland\\
$^3$Mastercard, Ireland\\
\{firstname.lastname\}@mastercard.com}\\
}

\maketitle

\maketitle
\begin{abstract}

While Large Language Models (LLMs) excel at tool-calling, deploying these capabilities in regulated enterprise environments like fintech presents unique challenges due to on-premises constraints, regulatory compliance requirements, and the need to disambiguate large, functionally overlapping toolsets. In this paper, we present a comprehensive study of tool retrieval methods for enterprise environments through the development and deployment of ScaleCall, a prototype tool-calling framework within Mastercard designed for orchestrating internal APIs and automating data engineering workflows. We systematically evaluate embedding-based retrieval, prompt-based listwise ranking, and hybrid approaches, revealing that method effectiveness depends heavily on domain-specific factors rather than inherent algorithmic superiority. Through empirical investigation on enterprise-derived benchmarks, we find that embedding-based methods offer superior latency for large tool repositories, while listwise ranking provides better disambiguation for overlapping functionalities, with hybrid approaches showing promise in specific contexts. We integrate our findings into ScaleCall's flexible architecture and validate the framework through real-world deployment in Mastercard's regulated environment. Our work provides practical insights into the trade-offs between retrieval accuracy, computational efficiency, and operational requirements, contributing to the understanding of tool-calling system design for enterprise applications in regulated industries.

\end{abstract}

\section{Introduction}

Tool calling has emerged as a key capability of large language models (LLMs), enabling them to interpret natural language instructions and dynamically invoke both internal and external application programming interfaces (APIs) or computational functions~\cite{schick2023toolformer, yao2023react, qin2023tool, patil2023gorilla}. This expands the role of LLMs from passive information retrievers to active agents capable of orchestrating workflows, executing operations, and integrating with real-world software systems. Recent research has demonstrated the promise of tool calling in general-purpose applications such as code generation~\cite{qin2023tool}, web automation~\cite{yao2023react}, and task decomposition~\cite{schick2023toolformer}, particularly in open-domain settings involving cloud-based or public APIs.

However, deploying tool-calling agents in regulated enterprise settings, such as financial services, presents unique challenges~\cite{schuman2025ai_pilots}. These environments feature complex internal toolchains composed of hundreds of proprietary services, including microservices and externally proxied APIs. Such tools often lack standardized documentation, exhibit overlapping naming conventions, and are governed by strict access controls. Regulatory compliance further imposes constraints around data handling, auditability, and system security. These issues are magnified in on-premises deployments, where agents must operate without access to external services or cloud-native tooling.

To better understand these challenges and explore practical solutions, we embarked on developing \textit{ScaleCall}, a tool-calling prototype framework within Mastercard designed to orchestrate internal APIs and automate data engineering tasks. Through this implementation effort, we conducted a comprehensive study of existing tool retrieval methods and their applicability to enterprise environments.

Current literature presents two primary approaches to tool retrieval: embedding-based similarity search and prompt-based listwise ranking. Embedding-based methods~\cite{qin2023tool, patil2023gorilla} offer fast retrieval by computing semantic similarity between queries and tool descriptions, making them suitable for latency-sensitive applications. However, as demonstrated by recent benchmarks such as ToolRet~\cite{shi2025retsavvy}, these approaches struggle with disambiguation when tools have overlapping functionalities. Conversely, prompt-based listwise ranking methods~\cite{sun2024chatgptgoodsearchinvestigating, zhang2023rankgpt} leverage LLMs' reasoning capabilities to perform comparative ranking, showing superior performance in domain-specific scenarios but at higher computational cost.

Through our empirical investigation within the ScaleCall framework, we studied both approaches individually and explored hybrid combinations. Our findings reveal that neither approach demonstrates consistent superiority across all scenarios. Instead, the effectiveness depends heavily on domain-specific factors such as tool repository size, semantic overlap between functions, and the nature of user queries. This observation motivated our design of a flexible retrieval architecture that can adapt to different enterprise contexts.

Consider, for instance, a Mastercard engineer automating account aggregation via Mastercard's Open Banking APIs~\cite{mastercard_openbanking}. This task requires coordination with internal consent management services that may expose similarly named endpoints (e.g., \texttt{getConsentDetails}) but differ in authorization logic, jurisdictional constraints, or intended business contexts. Without precise disambiguation, an LLM agent risks invoking the wrong service—leading to silent failures or policy violations. These scenarios demand domain-aware retrieval, execution gating, and permission validation—capabilities that must be tailored to specific enterprise requirements.

Emerging standards such as the \textit{Model Context Protocol} (MCP)~\cite{anthropic2024mcp, modelcontextprotocol2024} aim to simplify tool integration by providing structured interfaces between LLMs and external systems. However, their practical adoption in large-scale enterprise environments remains limited due to legacy systems, jurisdiction-specific compliance requirements, and fragmented tool repositories that complicate integration within real-world infrastructure.

Based on our observations from implementing ScaleCall, we identify four key dimensions that characterize enterprise tool-calling challenges: scalability (handling hundreds of overlapping tools), retrieval precision (disambiguating functionally similar services), deployment constraints (on-premises requirements and regulatory compliance), and functional scope (supporting robust execution with permission validation and error recovery).

In this work, we present our comprehensive study of tool retrieval methods for enterprise environments, conducted through the development and deployment of ScaleCall. Our investigation encompasses both traditional embedding-based approaches and modern listwise ranking techniques, as well as hybrid combinations of these methods. We evaluate these approaches using benchmarks derived from real-world enterprise use cases and provide insights into their practical trade-offs in regulated financial services environments.

To ensure comprehensive evaluation, we introduce a domain-aware assessment strategy that combines automated syntactic checks using ChatGPT\footnote{OpenAI's ChatGPT model, a large-scale generative language model} with manual validation by subject matter experts. This methodology enables scalable yet contextually grounded evaluation of tool selection quality in realistic enterprise scenarios.

\subsection*{\textbf{Key contributions:}}

\begin{itemize}
    \item \textbf{Enterprise tool-calling framework:} We present ScaleCall, a prototype tool-calling framework developed within Mastercard for orchestrating internal feature store APIs and automating data engineering workflows in regulated environments.

    \item \textbf{Comprehensive method comparison:} Through systematic experimentation, we evaluate embedding-based retrieval, prompt-based listwise ranking, and hybrid approaches, revealing that effectiveness depends on domain-specific factors rather than inherent method superiority.

    \item \textbf{Domain-aware evaluation methodology:} We design an evaluation framework that combines synthetic query generation, automated validation, and expert review to assess tool retrieval performance in enterprise contexts.

    \item \textbf{Deployment insights for fintech:} We provide empirical findings from real-world implementation, highlighting practical considerations for deploying tool-calling systems in regulated financial services environments, including performance trade-offs and operational requirements.
\end{itemize}
\section{Related Work}

\subsection{LLM-Based Tool Calling Paradigms}

Research on tool-calling has gained significant momentum in recent years \cite{qin2023toolllmfacdd, patil2023gorilla, yao2023reactsyting}, driven by the need to augment large language models with external capabilities. Current approaches typically follow two primary paradigms: retrieval-augmented generation (RAG) and instruction-following LLMs with in-context execution capabilities \cite{gao2023retrieval, fan2024survey}. RAG-based systems retrieve relevant tools based on semantic similarity and condition the LLM with their descriptions to generate function calls, while instruction-tuned models embed tool context within prompts and rely on the model's reasoning capabilities. High-level orchestration frameworks like HuggingGPT \cite{shen2023hugginggpt} demonstrate the potential of using LLMs as controllers to select and coordinate specialized models. However, most implementations are optimized for cloud-based, open-domain environments and fall short of meeting the stringent requirements of regulated sectors such as finance and healthcare \cite{schuman2025ai_pilots}. Recent work on edge deployment \cite{erdogan2024tinyagent} addresses computational constraints but does not tackle the unique challenges of enterprise environments.

\subsection{Tool Retrieval and Disambiguation}
The tool retrieval phase involves selecting appropriate APIs or functions for given queries. Foundational methods employ embedding-based similarity search over tool descriptions \cite{qin2023toolllmfacdd}, as implemented in frameworks like LangChain and OpenAI's function calling interface. To address retrieval ambiguity, recent works introduce sophisticated mechanisms: ToolLLM employs LLM-based entailment re-ranking, while query reformulation approaches \cite{kachuee2024improving} use LLMs to enhance embedding-based search effectiveness. The ToolRet benchmark \cite{shi2025retsavvy} systematically evaluates existing information retrieval models on tool retrieval tasks, revealing that even models with strong performance on conventional IR benchmarks exhibit poor performance when selecting from large, functionally overlapping toolsets. This benchmark demonstrates the critical need for domain-specific tool disambiguation, particularly in enterprise environments where repositories often contain hundreds of overlapping functions with minimal documentation and domain-specific naming conventions.

\subsection{Re-ranking Mechanisms for Tool Selection}
To address retrieval ambiguity, several re-ranking paradigms have emerged. Traditional approaches employ pointwise cross-encoders that score each (query, tool) pair in isolation. More recent work leverages general-purpose LLMs as re-rankers through structured frameworks like RankGPT \cite{sun2024chatgptgoodsearchinvestigating}, which uses permutation or sliding-window algorithms to manage long candidate lists. An emerging trend focuses on direct, single-pass listwise re-ranking \cite{zhang2023rankwithoutgpt}, where LLMs perform comparative ranking of all candidates simultaneously. This approach enables richer contextual reasoning compared to pairwise methods, making it particularly suitable for disambiguating tools with overlapping functionalities—a common challenge in enterprise environments.

\subsection{Enterprise AI and LLM Orchestration}
Enterprise deployment of LLM-powered systems presents unique challenges including security constraints, regulatory compliance, and integration with legacy infrastructure \cite{ibm2024orchestration, pizurica2024evaluating}. Recent industry trends show LLMs acting as orchestration layers that translate natural language requests into multiple API calls \cite{dibia2024agents}, minimizing manual intervention in complex workflows. However, enterprise environments require specialized considerations: secure sandboxing, permission validation, robust error handling, and integration with existing access control systems \cite{nvidia2024llm, orq2024llmops}. Agentic RAG systems for enterprise-scale deployment \cite{li2024agentic} begin to address these requirements, but comprehensive frameworks that handle the full tool-calling lifecycle in regulated environments remain limited.

\subsection{Challenges and Gaps}

Despite these advances, significant challenges persist in enterprise tool-calling systems. As highlighted by the ToolRet evaluation \cite{shi2025retsavvy}, even state-of-the-art retrievers fail to consistently select correct tools when presented with abstract or ambiguous task descriptions. This performance gap is particularly acute in enterprise settings where tools often lack unified documentation, display semantic overlap in naming conventions, and are governed by strict role-based access controls. The operational realities of enterprise usage—including the need for on-premises deployment, compliance with industry regulations, and integration with complex internal toolchains—underscore the importance of domain-specific reasoning, layered retrieval strategies, and comprehensive evaluation frameworks designed specifically for regulated environments. 
\section{Methodology}\label{agentic_design}

In this section, we outline the methodological steps undertaken in two distinct phases. First, we leverage the ToolRet benchmark~\cite{shi2025retsavvy}, including both its dataset and state-of-the-art retrieval strategies, to address the tool retrieval task, which is formalized as follows: given a user query (optionally accompanied by an \texttt{instruction}), the system must identify the most appropriate tool from a large candidate library. Second, we extend and adapt our implementation of these methods to accommodate the practical constraints of an enterprise-scale, on-premises deployment—specifically within our internal system, ScaleCall.

\subsection{ToolRet Reproduction Setup}
For this reproduction setup, we utilize the full ToolRet benchmark\footnote{\url{https://huggingface.co/mangopy/ToolRet-trained-e5-base-v2}, accessed June 2025}, a publicly available dataset specifically designed for tool retrieval in LLM-based systems. We adopt the dataset, the official evaluation protocol, and a modified version of the original retrieval strategy.

\subsubsection{Dataset and Evaluation}
As summarized in Table~\ref{tab:toolret_summary} (first column), ToolRet contains 7,615 retrieval tasks over a corpus of more than 43,000 tools, aggregated from over 30 sources including web APIs, software functions, and user-defined tools. Each task consists of a natural language query and instruction, tool metadata (name and description), and an annotated ground-truth tool ID. A representative example is shown in Table~\ref{tab:toolret_example}.

We analyzed the dataset version\footnote{Downloaded from Hugging Face in June 2025} used in our experiments and observed minor deviations in the number of tasks and tools compared to the originally reported statistics, as shown in Table~\ref{tab:toolret_summary}. Although the dataset documentation does not clarify the source of these discrepancies, possible explanations include version updates or differences in data preprocessing scripts. Our evaluation is based on the observed counts to ensure consistency with the specific version employed in our experiments.

\begin{table*}[htbp]
\centering
\scriptsize
\caption{Sample entry from the TOOLRET benchmark dataset. For our use case, the attributes of interest are \texttt{Query}, \texttt{Instruction}, \texttt{Tool ID}, and \texttt{Description}. The \texttt{Tool Name} field is used solely for tool mapping during retrieval.}
\begin{tabular}{p{3cm} p{10cm}}
\toprule
\textbf{Attribute} & \textbf{Example Value} \\
\midrule
Query & \makecell[{{p{10cm}}}]{How can I resize an image to specific dimensions?} \\
Instruction & \makecell[{{p{10cm}}}]{Use a tool that resizes an image to the specified width and height while maintaining the aspect ratio.} \\
Tool Name & \makecell[{{p{10cm}}}]{\texttt{resize\_image\_tool}} \\
Description & \makecell[{{p{10cm}}}]{Resizes a given image to the specified width and height. Maintains aspect ratio by default. Accepts input as an image URL or base64 string, and returns the resized image in base64 format.} \\
Category & \makecell[{{p{10cm}}}]{Web API} \\
Tool ID & \makecell[{{p{10cm}}}]{\texttt{tool\_4398}} \\
Domain & \makecell[{{p{10cm}}}]{Image Processing} \\
Input Schema & \makecell[{{p{10cm}}}]{\{ "image": "string", "width": "int", "height": "int" \}} \\
Output Schema & \makecell[{{p{10cm}}}]{\{ "resized\_image": "string" \}} \\
\bottomrule
\end{tabular}
\label{tab:toolret_example}
\end{table*}

To ensure the validity and generalizability of our results, we adopt the standard evaluation protocol recommended by the ToolRet benchmark, using the following retrieval metrics:
\begin{itemize}
\item \textbf{Success Rate (P@1):} Percentage of queries where the top-ranked tool is correct.

\item \textbf{Recall@k (R@k):} Percentage of queries where the correct tool appears in the top-$k$ results ($R@1$, $R@10$).

\item \textbf{Precision@k (P@k):} Proportion of relevant tools among the top-$k$ retrieved ($P@1$, $P@10$).

\item \textbf{Completeness@k (C@k):} Percentage of queries where all ground-truth tools appear in the top-$k$.

\item \textbf{nDCG@k (N@k):} Normalized Discounted Cumulative Gain, capturing ranking quality by rewarding higher-ranked relevant tools.
\end{itemize}

\begin{table*}[thbp]
\centering
\scriptsize
\setlength\tabcolsep{4pt}
\caption{ToolRet Dataset Statistics: Comparison between reported and observed values. The reported statistics are taken from the original release paper (ToolRet\cite{shi2025retsavvy}), while the observed statistics are derived from our own analysis of the benchmark, accessed via the Hugging Face GitHub repository on July 7, 2025. The "Source Datasets" count could not be verified as the provided data was pre-processed and aggregated, hence it is marked as (–) in our analysis.}

\begin{tabular}{p{5.2cm} p{2.4cm} p{2.8cm}}
\toprule
\textbf{Statistic} & \textbf{As Reported} & \textbf{Observed} \\
\midrule

\rowcolor{blue!5}
Tool Corpus Size & 43,000+ & 44453 \\
\rowcolor{blue!5}
Source Datasets & 30+ & -- \\

\# Retrieval Tasks & 7,615 & 7,961\\
\quad -- Web API Retrieval Tasks & 4,916 & 5,230\\
\quad -- Code Function Retrieval Tasks & 950 & 1,749\\
\quad -- Customized App Retrieval Tasks & 1,749 & 982 \\
\addlinespace
\# Tools & 43,215 & 44,453 \\
\quad -- Web APIs & 36,978 & 37,292 \\
\quad -- Code Functions & 3,794 & 3,794 \\
\quad -- Customized Apps & 2,443 & 3,367 \\


\addlinespace
Avg. Query / Instruction Length (tokens) & 46.87 / 43.43 & 34.98 / 35.67 \\
Avg. Tool Documentation Length (tokens) & 174.56 & 68.0\\
\bottomrule
\end{tabular}
\label{tab:toolret_summary}
\end{table*}

\subsubsection{Retrieval Strategies}

\begin{figure*}[htbp]

        \centering
        \includegraphics[width=0.4\textwidth]{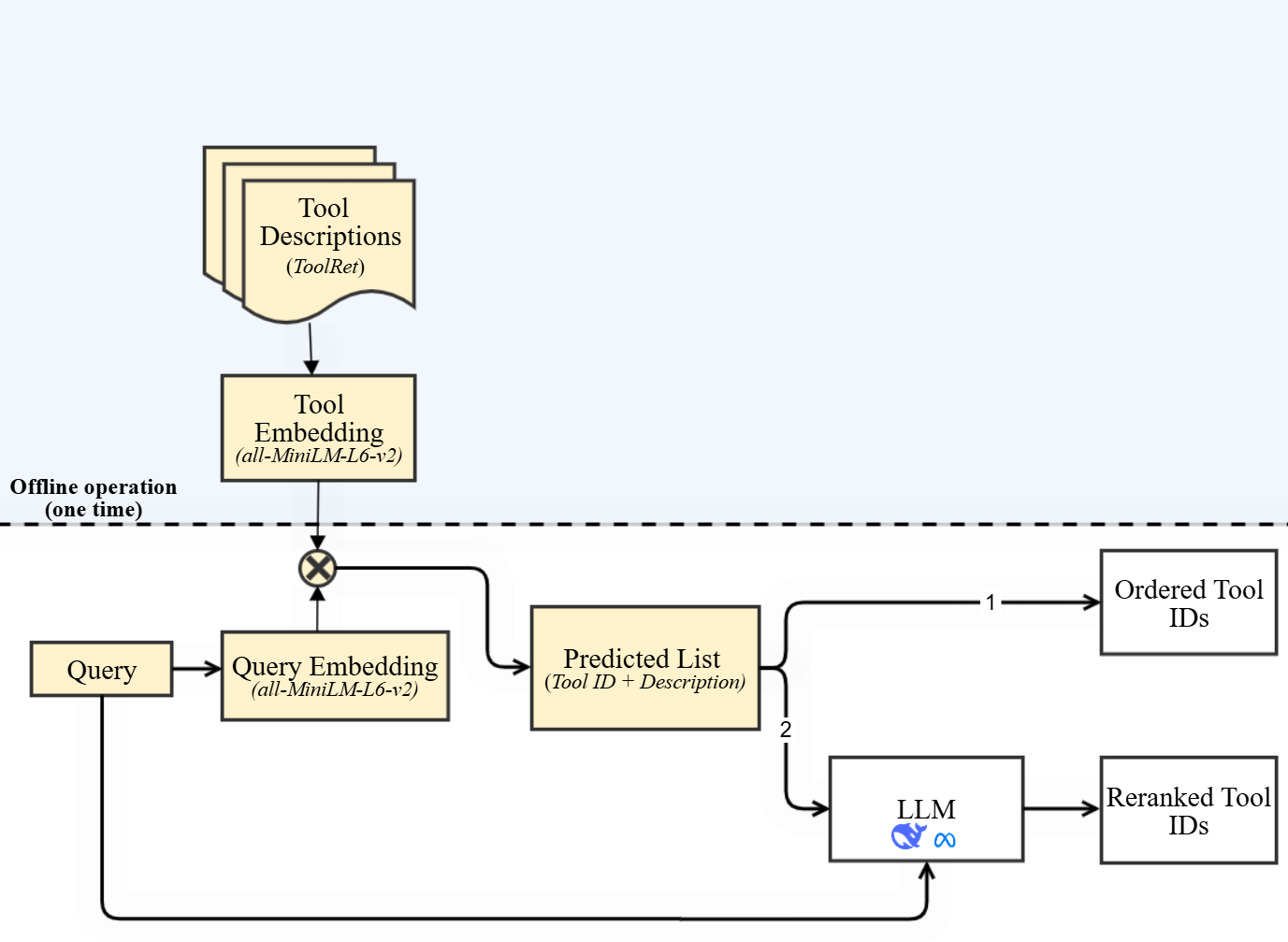} 
        \caption{Architecture of the Tool Retrieval Framework. The diagram illustrates both the baseline architecture, Embedding-based Tool Retriever (ETR), and the enhanced architecture incorporating the re-ranking module, referred to as Re-ranked Tool Retriever (RTR). Components highlighted in \colorbox{lightyellow}{yellow} indicate shared modules utilized by both approaches. Flow (1) corresponds to the ETR pipeline, which generates an ordered list of tools based on embedding similarity scores. Flow (2) represents the extension introduced in RTR, where a large language model (LLM) re-ranker refines the initial ranking. The symbol \textbf{\colorbox{lightyellow}{$\bm{\otimes}$}} denotes the embedding similarity operation, such as cosine similarity or dot product.}
        \label{fig:system_diagram}
\end{figure*}

We adopt two primary retrieval strategies from the ToolRet benchmark: (1) the baseline embedding-based retrieval and (2) a refined re-ranking architecture designed to improve precision and robustness. The overall system architecture is illustrated in Figure~\ref{fig:system_diagram}, where components highlighted in yellow denote shared modules across both methods. Flow (1) corresponds to the baseline retrieval approach, while Flow (2) represents the re-ranking pipeline.


\begin{figure*}[htbp]
        \centering
        \includegraphics[width=0.6\textwidth]{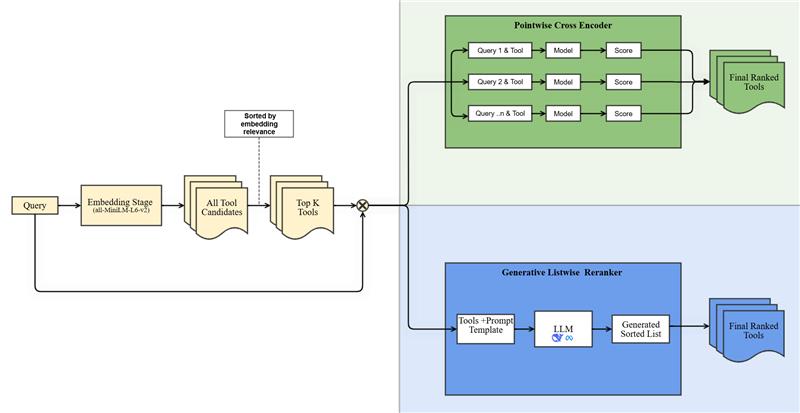} 
            \caption{A comparative illustration of two re-ranking architectures. The left side of the diagram (white background) depicts the shared initial embedding stage employed by both our systems (ETR and TTR), and by the original ToolRet framework~\cite{shi2025retsavvy} for retrieving the top-$k$ candidate tools. The divergence occurs in the re-ranking phase. ToolRet applies a Pointwise Cross-Encoder (top-right, \colorbox{myLightGreen}{green}), whereas our proposed method employs a Generative Listwise Re-ranker (bottom-right, \colorbox{myLightBlue}{blue}). The symbol \textbf{\colorbox{myLightYellow}{$\bm{\otimes}$}} indicates the point at which the user query is combined with the candidate tools for the re-ranking operation.}
            \label{fig:theirapp}
\end{figure*}

\paragraph{Baseline: Embedding-based Tool Retriever (ETR)}

Our baseline follows the baseline proposed in ToolRet. The ETR ranks tools based on the semantic similarity between the user query and tool descriptions, using static embeddings generated by a pre-trained sentence encoder. Tool embeddings are computed offline, while queries are encoded at runtime using the same model. Cosine similarity is then used to retrieve the top-$k$ most relevant tools. Flow (1) in Figure~\ref{fig:system_diagram} illustrated the flow of this pipeline. 

This method is computationally efficient and yields high recall. However, its reliance on shallow similarity metrics limits its ability to distinguish between semantically similar but functionally distinct tools, which reduces precision in practice.

\paragraph{Tool Retrieval with Re-ranking (TRR)}

To address the limitations of ETR, we implement an enhanced strategy termed \textit{Tool Retrieval with Re-ranking (TRR)}. In this approach, the list of candidate tools retrieved by ETR is passed to a large language model (LLM), which performs a second round of semantic evaluation. The LLM matches the embeddings of both the query and the candidate tools to produce a refined, final ranking. This process corresponds to Flow 2 in Figure~\ref{fig:system_diagram}.

Our re-ranking implementation differs architecturally from that of the original ToolRet design, as illustrated in Figure~\ref{fig:theirapp}. While ToolRet employs a pointwise re-ranking strategy using a cross-encoder to evaluate each candidate independently, we adopt a listwise re-ranking approach. In our setup, the LLM jointly reasons over the full set of top-$k$ candidates retrieved by the ETR stage, enabling comparative analysis across tools.

We argue that listwise re-ranking is better suited to this task, as it captures inter-tool relationships and functional subtleties that pointwise methods may overlook. This distinction is particularly important in enterprise environments, where precision and context-awareness in tool invocation are critical for operational reliability.

\subsection{Adaptation to Enterprise Constraints}
The replication of the ToolRet benchmark in this work is intended to serve as a research foundation for designing the tool retrieval component within the ScaleCall system. However, a full reproduction of ToolRet’s experimental setup was not feasible due to constraints inherent to enterprise deployment environments. Consequently, we introduced methodological adaptations to align with infrastructure and operational requirements.

One of the primary constraints relates to the scope of model experimentation. Unlike the broader suite of models evaluated in ToolRet, our implementation of the ETR and TRR approaches was restricted to a subset of models that met internal compliance criteria. Specifically, only models that were publicly available, vetted for operational use, and approved for deployment within our enterprise environment were considered. Factors influencing this selection included licensing limitations, cost constraints, and hardware requirements that precluded the on-premises deployment of certain larger models.

These adaptations ensured that the reproduced system remains consistent with organizational policies while providing actionable insights for integrating LLM-based tool retrieval in regulated fintech settings. 
\section{Experiment and results}

\subsection{Reproduction Validation}
\subsubsection{Experimental Setup}
We evaluate the ETR and TTR models using the ToolRet benchmark\footnote{\url{https://huggingface.co/mangopy/ToolRet-trained-e5-base-v2}, accessed June 2025}. The benchmark is employed in its default configuration, and evaluation is carried out using the official metrics outlined in Section~\ref{agentic_design}. Our analysis covers three specific subsets of the benchmark: \textit{Web}, \textit{Code}, and \textit{Customized—}, as well as the overall dataset.

To assess the role of prompt informativeness, we conduct evaluations under two input settings: (a) query only which simulates under-specified inputs (\texttt{query} $-$ \texttt{instruction}), and (b) query plus instruction which provides full context (\texttt{query} $+$ \texttt{instruction}). Query and tool embeddings are generated using \texttt{all-MiniLM-L6-v2}, and re-ranking in the TTR pipeline is performed using \texttt{LLaMA 3.1-8B} and \texttt{LLaMA 3.2-3B}, served via vLLM. The re-ranker reorders the top-$k$ candidates ($k=10$) from the initial retrieval stage. The experimentation in this study is restricted to the \texttt{all-MiniLM-L6-v2}, \texttt{LLaMA 3.1-8B}, and \texttt{LLaMA 3.2-3B} models due to their suitability for enterprise deployment. These models are publicly available, offer a favorable balance between model size and reasoning capabilities, and can be efficiently deployed in on-premise environments. Most importantly, they have been vetted and approved for use within our regulated operational setting. 

Another operational adaptation we applied was standardizing all embedding computations to use \texttt{float32} precision for Layer Normalization on CPUs, instead of the default \texttt{float16}. This decision was made to ensure numerical stability and maintain consistent behavior across varying hardware configurations.

\subsubsection{Results}

Tables~\ref{tab:results_wo_inst_small} and~\ref{tab:results_with_inst_small2} present the performance results under the "without instruction" and "with instruction" settings, respectively. Our replication of the baseline (ETR) aligns closely with the original ToolRet results reported by Shi et al.~\cite{shi2025retsavvy}, with only minor deviations. These differences are primarily attributed to two factors: (1) the use of an updated benchmark version containing a slightly higher number of tasks (7,961 vs. 7,615), particularly an increase in TOOLRET-Code instances, which affects the overall aggregated metrics; and (2) the use of \texttt{float32} precision for embedding computations, rather than the \texttt{float16} used in the original setup. Despite these variations, the consistency of the observed results with those of the original benchmark confirms the robustness and fidelity of our baseline reproduction.

In contrast, TRR performance diverged due to enterprise-specific constraints. While the re-ranker showed strong results under lab conditions, on-premise deployment led to frequent timeouts, high latency, and reduced re-ranking quality. These operational challenges significantly diminished the practical benefits of re-ranking in constrained environments.

\paragraph{Impact of Instructions}

A key finding from the experimental results is the consistent and substantial performance improvement achieved through the inclusion of instructions across all datasets and models. Comparing Table~\ref{tab:results_wo_inst_small} and Table~\ref{tab:results_with_inst_small2}, we observe notable gains for both the baseline (ETR) and the re-ranking (TRR) setups. For ETR, the average $N@10$ increased from 14.26\% to 32.38\% on the TOOLRET-Code subset and from 24.78\% to 33.29\% on TOOLRET-Custom. In the TRR setting, particularly with stronger models such as \texttt{LLaMA 3.1-8B}, the average $N@10$ improved from 13.82\% to 15.80\%. The most pronounced uplift was observed on the TOOLRET-Web subset, where $N@10$ rose from 14.35\% to 30.35\% for \texttt{LLaMA 3.2-3B}, and from 15.92\% to 34.90\% for \texttt{LLaMA 3.1-8B}. Similar improvements were seen on TOOLRET-Custom, with increases from 22.73\% to 28.00\% and from 28.00\% to 35.48\% for \texttt{LLaMA 3.2-3B} and \texttt{LLaMA 3.1-8B}, respectively. These results highlight the critical role of instruction-based prompting in improving retrieval effectiveness, particularly when using advanced language models in re-ranking scenarios.

\paragraph{Performance Across Datasets}
Performance varies across the three ToolRet subsets. The \textit{TOOLRET-Web} subset yields the highest overall scores and shows the most significant improvement with the inclusion of instructions. In contrast, \textit{TOOLRET-Code} proves to be the most challenging, with consistently lower $N@10$ and $Completeness@10$ scores across models. This is likely due to the technical nature of the queries. The \textit{TOOLRET-Customized} subset, which includes complex and user-defined tools, demonstrates the strong capability of the \texttt{LLaMA 3.1-8B} model which indicates its effectiveness in handling more nuanced retrieval tasks.

\subsection{Enterprise Insights}

Given the operational focus of our experimentation in enterprise settings, we present below the key observations from an infrastructure-aware deployment perspective. The deployment of either the embedding-based baseline approach (ETR) or the hybrid retrieval-and-re-ranking approach (TRR) exhibits significant sensitivity to enterprise infrastructure constraints.
\begin{itemize}
    \item Model selection was dictated not solely by performance metrics, but also by compliance with enterprise requirements. Although the ToolRet benchmark evaluates a broader suite of models—some of which outperform \texttt{all-MiniLM-L6-v2}, \texttt{LLaMA 3.1-8B}, and \texttt{LLaMA 3.2-3B}—these higher-performing models could not be adopted in our context due to licensing restrictions, cost considerations, or hardware limitations preventing their on-premises deployment. As such, our experimental scope was limited to models that were publicly available, operationally vetted, and approved for use within our regulatory environment.

    \item Infrastructure sensitivity emerged as a critical factor, particularly for the LLM-based re-ranking component. Even minor latency spikes in the generative re-ranking step frequently triggered fallback mechanisms, thus constraining the applicability of TRR in latency-sensitive workflows.

    \item Embedding-based models demonstrated strong trade-offs between efficiency and performance. The ETR approach sustained acceptable levels of precision and recall while maintaining low latency, making it well-suited for production systems where responsiveness is paramount.

    \item Prompt enrichment, especially the inclusion of task-specific instructions, yielded substantial gains in retrieval effectiveness. In some subsets, the N@10 metric improved by as much as 77.6\%, underscoring the importance of high-quality input formulation for maximizing model utility.

\end{itemize}

Collectively, these insights highlight a crucial principle for enterprise deployment: in regulated, resource-constrained environments, system simplicity and thoughtful input design frequently offer more operational value than increased architectural complexity.

\begin{table*}[t]
\centering
\scriptsize
\caption{Performance Comparison Without Instructions for ETR and TRR. This table presents retrieval results for a low-context scenario where models receive only the raw user query. Colors indicate: \textcolor{ours}{blue} for our results, \textcolor{reported}{gray*} for the reported baseline from the original ToolRet benchmark, and \textcolor{reranker}{green} for models incorporating re-ranking.}
\label{tab:results_wo_inst_small}
\begin{tabular}{l|cccc|cccc|cccc|cc}
\toprule
\multirow{1}{*}{Model} & \multicolumn{4}{c|}{Web} & \multicolumn{4}{c|}{Code} & \multicolumn{4}{c|}{Custom} & \multicolumn{2}{c}{Avg.} \\
 & N@10 & P@10 & R@10 & C@10 & N@10 & P@10 & R@10 & C@10 & N@10 & P@10 & R@10 & C@10 & N@10 & C@10 \\
\midrule
\textcolor{reported}{all-MiniLM-L6-v2$^*$} & 11.66 & 3.07 & 16.36 & 10.15 & 14.44 & 2.50 & 19.50 & 18.11 & 22.80 & 5.21 & 29.10 & 20.25 & 16.30 & 16.17 \\
\midrule
\textcolor{ours}{all-MiniLM-L6-v2 (ours)} & 11.71 & 2.71 & 16.41 & 11.36 & 14.26 & 2.39 & 18.69 & 17.47 & 24.78 & 5.71 & 32.65 & 24.45 & 16.92 & 17.76 \\
\midrule

\textcolor{reranker}{LLaMA 3.2-3B} & 11.97 & 2.71 & 16.41 & 11.36 & 14.35 & 2.39 & 18.69 & 17.47 & 22.73 & 5.71 & 32.65 & 24.45 & 16.35 & 17.76 \\
\textcolor{reranker}{LLaMA 3.1-8B}  & \textbf{13.82} & \textbf{2.71} & \textbf{16.41} & \textbf{11.36} & \textbf{15.92} & \textbf{2.39} & \textbf{18.69} & \textbf{17.47} & \textbf{28.00} & \textbf{5.71} & \textbf{32.65} & \textbf{24.45} & \textbf{19.25} & \textbf{17.76} \\

\bottomrule
\end{tabular}

\end{table*}

\begin{table*}[t]
\centering
\scriptsize
\caption{Performance comparison with instructions. This table presents retrieval results for a high-context scenario where models receive both the user query and its associated instruction. Colors indicate: \textcolor{ours}{blue} for our results, \textcolor{reported}{gray*} for the reported baseline from the original ToolRet benchmark, and \textcolor{reranker}{green} for models incorporating embedding and re-ranking.}
\label{tab:results_with_inst_small2}
\begin{tabular}{l|cccc|cccc|cccc|cc}
\toprule
\textbf{Model} & \multicolumn{4}{c|}{\textbf{Web}} & \multicolumn{4}{c|}{\textbf{Code}} & \multicolumn{4}{c|}{\textbf{Custom}} & \multicolumn{2}{c}{\textbf{Avg.}} \\
 & N@10 & P@10 & R@10 & C@10 & N@10 & P@10 & R@10 & C@10 & N@10 & P@10 & R@10 & C@10 & N@10 & C@10 \\
\midrule
\textcolor{reported}{all-MiniLM-L6-v2$^*$} & 12.77 & 3.26 & 19.38 & 13.33 & 31.59 & 5.06 & 43.86 & 42.25 & 32.24 & 7.14 & 43.55 & 32.34 & 25.53 & 29.31 \\
\midrule
\textcolor{ours}{all-MiniLM-L6-v2 (ours)} & 13.24 & 3.02 & 19.49 & 14.24 & 32.38 & 5.06 & 43.93 & 42.33 & 33.29 & 7.18 & 43.95 & 32.41 & 26.30 & 29.66 \\
\midrule

\textcolor{reranker}{Llama 3.2-3B} & 13.02 & 3.01 & 19.46 & 14.24 & 30.35 & 5.06 & 43.93 & 42.33 & 28.08 & 7.18 & 43.95 & 32.41 & 23.82 & 29.66 \\
\textcolor{reranker}{Llama 3.1-8B} & \textbf{15.80} & \textbf{3.00} & \textbf{19.35} & \textbf{14.13} & \textbf{34.90} & \textbf{5.06} & \textbf{43.93} & \textbf{42.33} & \textbf{35.48} & \textbf{7.18} & \textbf{43.95} & \textbf{32.41} & \textbf{28.72} & \textbf{29.62} \\

\bottomrule
\end{tabular}

\end{table*}

\section{System Integration: ScaleCall}
\begin{figure*}
        \centering
        \includegraphics[width=0.5\textwidth]{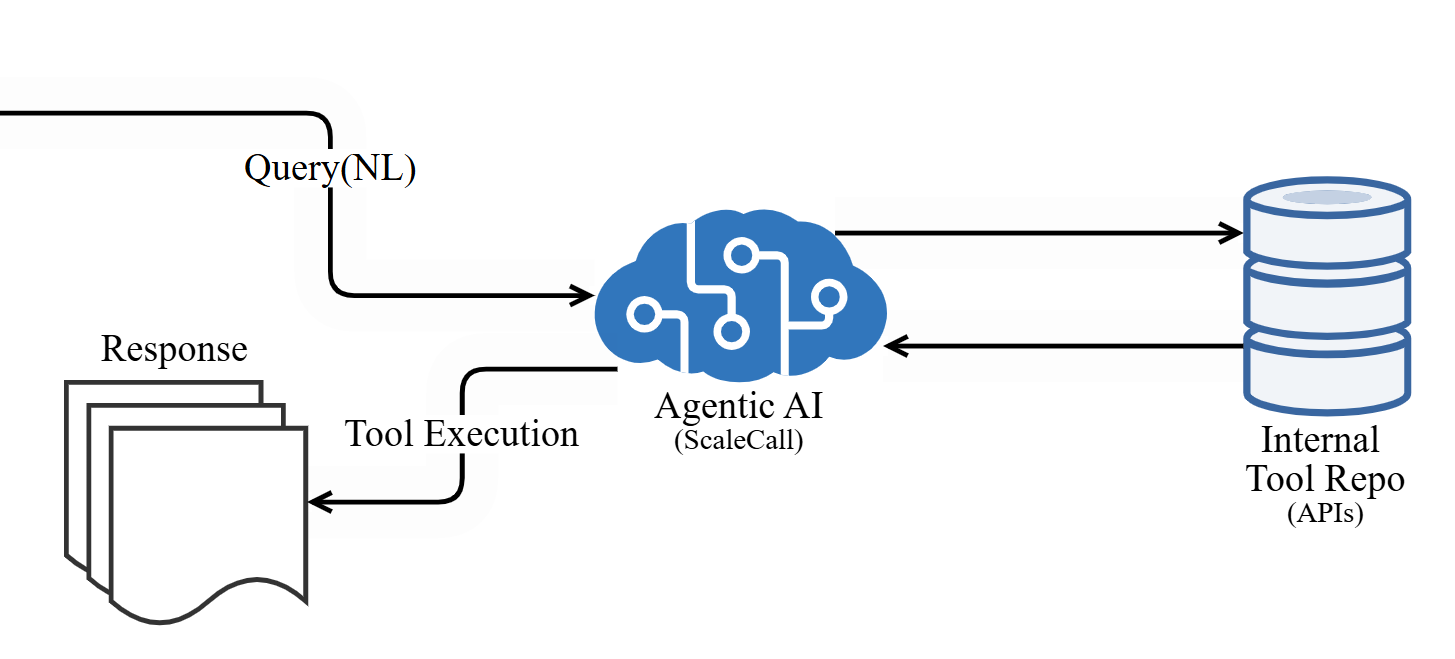} 
        \caption{High-level Diagram of ScaleCall Architecture}
        \label{fig:scalecall_diagram}
\end{figure*}

We leveraged insights gained from replicating the ToolRet experiments to inform the development of ScaleCall, an internal tool retrieval system deployed within the Mastercard environment. An extended version of the Tool Retrieval with Re-ranking (TRR) approach was adapted and integrated into enterprise workflows to support tool invocation tasks via a natural language interface. To broaden its applicability across diverse stakeholder use cases, we augmented the framework with Natural Language Inference (NLI) and NLI-based querying capabilities. These enhancements were incorporated into two proprietary tool repositories, enabling more fine-grained interpretation of user intent and improving classification accuracy by leveraging entailment and contradiction signals. This real-world deployment illustrates the method’s adaptability to enterprise-scale settings and underscores the added value of NLI integration for enhancing contextual understanding. A high-level overview of the system architecture is shown in Figure~\ref{fig:scalecall_diagram}.

\section{Discussion}

\paragraph{The Primacy of Contextual Input over Model Architecture}
One of the most prominent insights from our study is that improvements in tool retrieval performance are more significantly driven by the inclusion of contextual information—particularly user-provided \texttt{instructions}—than by the choice of underlying model architecture. For instance, the top-performing model, \texttt{LLaMA 3.1-8B}, demonstrated marked gains in $N@10$ scores on both the TOOLRET-Code and TOOLRET-Custom subsets when instructions were appended to the input query. This observation suggests that the expressiveness and clarity of the prompt, especially in articulating user intent and task constraints - shows a stronger influence on retrieval accuracy than architectural differences across models. These findings highlight the critical role of prompt engineering in retrieval-oriented tasks. In real-world deployments, particularly those involving LLM-powered agentic workflows, introducing a pre-processing stage aimed at augmenting raw user queries with structured intent or contextual metadata may yield significant accuracy gains, independent of the retrieval architecture employed.

\paragraph{The Effect of Model Scale and the Surprising Performance Plateau}

As anticipated, the larger \texttt{LLaMA 3.1-8B} model consistently outperforms its smaller counterpart, \texttt{LLaMA 3.2-3B}, reinforcing the established relationship between model scale and retrieval performance. However, a surprising pattern emerges when comparing \texttt{LLaMA 3.1-8B} to the specialized embedding model \texttt{all-MiniLM-L6-v2}. In multiple evaluation settings, including the “with instruction” configuration on the TOOLRET-Web subset, the performance metrics of both models are nearly identical—\texttt{LLaMA 3.1-8B} achieves an $N@10$ of 13.24\% and a $P@10$ of 3.02, while \texttt{all-MiniLM-L6-v2} records 13.02\% and 3.00 for the same metrics, respectively.

This convergence is not a measurement anomaly but a direct outcome of our fallback mechanism. To ensure reliability under production constraints, the system defaults to the embedding-based re-ranker whenever latency from serving complex listwise prompts through vLLM exceeds the predefined timeout thresholds. While this design improves system robustness, it highlights a broader engineering challenge: the performance of LLM-based pipelines can be bottlenecked by latency-sensitive components, regardless of the underlying model’s capacity.

At the same time, these results reinforce the strength of the \texttt{all-MiniLM-L6-v2} model. Despite its relatively small size and computational efficiency, it delivers competitive retrieval performance when compared to significantly larger language models. This underscores the practical utility of optimized embedding models, particularly in environments where inference speed and scalability are critical.

\paragraph{Limitations of the Current Study}
A key limitation of our TRR experiments was the operational instability of the LLM-based re-ranker. Re-ranking large candidate sets led to long, compute-intensive prompts that often exceeded latency thresholds, triggering our fallback to the embedding-based (ETR). Consequently, much of the reported TRR performance reflects the baseline model rather than the intended LLM-based component.
In addition to infrastructure-related constraints, our study was limited by model selection restrictions imposed by our enterprise environment. Specifically, we were required to use only pre-approved foundation models for deployment. Although these models, such as \texttt{LLaMA 3.1-8B} and \texttt{LLaMA 3.2-3B},exhibited strong performance, prior work, including the original TOOLRET study, reported superior results using alternative models. Consequently, our evaluation does not reflect the full range of potential performance achievable with more permissive model choices. Future iterations of this work should revisit the model selection process, potentially incorporating more diverse or specialized architectures that have been empirically shown to perform better in tool retrieval tasks. 
\section{Conclusion}
In this paper, we presented ScaleCall, a tool-calling framework developed for deployment within Mastercard’s regulated, on-premises environment. Through systematic evaluation of embedding-based, prompt-based, and hybrid retrieval methods on enterprise-relevant benchmarks, we identified critical trade-offs between retrieval accuracy, latency, and contextual disambiguation. Our findings emphasize that effectiveness is less a matter of algorithmic superiority and more dependent on input structure and domain-specific constraints. Notably, the inclusion of instructional context led to substantial performance gains, affirming the importance of query enrichment in real-world agentic workflows. We observed that the baseline embedding models offer strong baseline recall with low latency and the  listwise re-ranking via LLMs introduces engineering challenges, particularly around latency with limited improvements. ScaleCall’s flexible architecture integrates these insights and demonstrates their viability through deployment in a live enterprise setting. Overall, our work offers a practical roadmap for designing robust, efficient, and compliant tool-calling systems in regulated industries.

\bibliographystyle{IEEEtran}
\bibliography{references} 

\end{document}